# Ice : The paradigm of wild plasticity

**Jérôme Weiss[1,*]**

[1]Univ. Grenoble Alpes, CNRS, ISTerre, 38000 Grenoble, France

**Keywords:** ice, plasticity, dislocation avalanches, fluctuations, anisotropy, strain hardening

## Summary

Ice plasticity has been thoroughly studied, owing to its importance in glaciers and ice sheets dynamics. In particular, its anisotropy (easy basal slip) has been suspected for a long time, then fully characterized 40 years ago. More recently emerged the interest of ice as a model material to study some fundamental aspects of crystalline plasticity. An example is the nature of plastic fluctuations and collective dislocation dynamics. 20 years ago, acoustic emission measurements performed during the deformation of ice single crystals revealed that plastic "flow" proceeds through intermittent dislocation avalanches, power law distributed in size and energy. This means that most of ice plasticity takes place through few, very large avalanches, thus qualifying associated plastic fluctuations as "wild". This launched an intense research activity on plastic intermittency in the Material Science community. The interest of ice in this debate is reviewed, from a comparison with other crystalline materials. In this context, ice appears as an extreme case of plastic intermittency, characterized by scale-free fluctuations, complex space and time correlations as well as avalanche triggering. In other words, ice can be considered as the paradigm of wild plasticity.

*Author for correspondence (jerome.weiss@univ-grenoble-alpes.fr).

# Main Text

1.    Introduction

The fascination that icy objects (glaciers, ice-sheets, sea ice) provoked to mankind for millennia turned into scientific curiosity by the end of the XVIIIth century when de Saussure climbed the Mont Blanc to discover and analyze glaciers and altitude[1]. Very soon in the history of dawning glaciology, the idea of a viscous flow of glacier ice was proposed by Rendu [2], then more formally studied by Agassiz, Forbes and Tyndall in the mid- XIXth century[3]. By the end of the same century, evidence of easy basal slip, i.e. of a strong viscoplastic anisotropy, was first shown by McConnel[4]. This anisotropy, which was later confirmed and quantified by many authors (e.g. [5-7]), is related to the hexagonal structure of ice Ih and is the essential characteristic of ice viscoplasticity. We will extensively discuss below how it contrasts with multi-slip plasticity of face-centered cubic (FCC) and body-centered cubic (BCC) metallic materials, leading to a fundamentally different collective dislocation dynamics and a different nature of the plastic "flow" in terms of intermittency, strain localization, strain-hardening, or internal sub-structure.

This plastic anisotropy has also important consequences for polycrystalline ice (e.g. glacier ice) plasticity: it generates strong strain incompatibilities between neighboring grains [8], which can be relaxed e.g. through dynamic recrystallization mechanisms [9-12]. These mechanisms have been extensively studied over the last decades, together with fabrics development[13], owing to their importance for glacier and ice-sheet flow modeling[14, 15], hence, in this last case, for ice cores dating (e.g. [15]). A detailed review can be found in [16]. Here we do not focus on these topics. Instead, we revisit some fundamental aspects of ice single crystal plasticity, with a brief incursion on polycrystals, in light of "recent" (the last 20 years) developments on the small scale (from the individual dislocation scale to the crystal/sample scale) dynamics of plastic deformation. We show that ice plastic "flow" actually occurs in a strongly spatially heterogeneous and intermittent way, through dislocation avalanches spanning a huge range of size. We relate this with the plastic anisotropy mentioned above that maximizes the role of long-range elastic interactions in collective dislocation dynamics, and discuss the link with the absence of significant isotropic hardening in ice or of development of an internal characteristic microstructural length scale. This "wild" dynamics, characterized by power law distributions of plastic fluctuations (with infinite variance)[17], strongly contrasts with the plasticity of FCC and BCC metals characterized, for sample sizes larger than few μm, by "mild" (finite variance, Gaussian-like) fluctuations, the development of dislocation sub-structures (e.g. dislocation cells) and isotropic hardening. In this context, ice appears as the paradigm for wild plasticity, including at bulk (>mm) scales.

2.      Some specificities of ice plasticity

The fundamentals of ice plasticity have been extensively detailed elsewhere [16, 18, 19]. Here, the goal is not to give such overview, but instead to focus on some specificities of ice plasticity that are playing a key role in collective dislocation dynamics.

*Plastic anisotropy*

As already stressed above, plastic anisotropy is the most prominent characteristic of ice plasticity. This has been exemplified by the creep tests reported by Duval and others [6] showing, for a given resolved shear stress, basal slip creep rates orders of magnitude larger than non-basal slip creep rates. It is related to the hexagonal structure of ice Ih, shared with hexagonal close packed (HCP) metals such as Zn of Cd, although particularly pronounced for Ice.

This slip anisotropy at the crystal/sample scale does not mean that non-basal dislocation motion is impossible [20]. The development of short segments of edge dislocations on prismatic planes have been observed [21, 22]. These non-basal segments, however, account for only a very limited fraction of macroscopic plasticity, as their extension is strongly limited. On the other hand, they are generally associated with the climb [23] or the cross-slip [24, 25] of basal dislocations. These configurations represent weak pinning points for individual dislocation motion, and consequently can play a role on dislocation sources. As an example, double cross-slip of basal screw dislocations might be an efficient mechanism of dislocation multiplication, and therefore of strain heterogeneity and localization[24-26]. The local pinning effect of these configurations are however likely much weaker than short-range interactions such as junctions between dislocations gliding on different slip systems, which give rise to forest hardening in the much more isotropic plasticity of FCC metals[27].

Counter-intuitively, the mobility (dislocation velocity/resolved shear stress) of individual non-basal edge dislocations is almost one order of magnitude larger than the mobility of basal ones[28]. This stresses the fundamental difference between individual dislocation dynamics in one hand, and collective dislocation dynamics in the other hand.

*Lattice resistance*

Lattice resistance in ice, $\tau_l$, which is associated to O-O bonds [29], is not yet known, but is likely very small, at least at temperatures relevant to lab experiments or terrestrial ice ($T > -50°C$). Laboratory mechanical tests can only give an upper bound: torsion creep tests performed at -10°C on well-oriented ice single crystals revealed a significant viscoplastic deformation through basal glide for a resolved shear stress as low as 30 kPa [24]. This would correspond to a normalized lattice resistance $\tau_l/G$ , where $G$ is the shear modulus, lower than $10^{-5}$. In addition, the shear strain-rate measured during these torsion tests under 30 kPa are larger than $10^{-8}$ s$^{-1}$, while ice sheet flow rates as low as $10^{-13}$-$10^{-12}$ s$^{-1}$ can be deduced from the ratio between the snow accumulation rate and the ice sheet thickness [30-32]. This suggests that $\frac{\tau_I}{G}$ might indeed be much smaller than $10^{-5}$.

*Strain-hardening*

In FCC metals, the multi-slip plasticity generates complex dislocation entanglements "stiffened" by short-range interactions between dislocations gliding on different planes. The resulting strain-hardening, which has been used for millennia to harden metals and alloys from forging, is isotropic, meaning that most of plastic strain is not recoverable upon unloading. Instead, strain-hardening in ice is essentially kinematic (or directional)[6]: while in single crystals strain-hardening is absent[33], it is known for a long time that a large part of plastic deformation is recoverable in polycrystals [34, 35], in relation with the relaxation of polarized internal stresses associated with dislocation arrangements such as dislocation pile-ups at grain boundaries, as long as recrystallization processes remains limited. Under creep (constant stress) loading conditions, kinematic hardening is associated with the primary creep regime, and for associated transient strains up to ~$10^{-4}$, most of the deformation is recoverable [35]. For much larger applied strains, up to ~$10^{-2}$ but still in the macroscopic primary creep regime, an increasing part of the strain becomes irreversible, in association with sub-boundaries formation, but the recoverable deformation remains important (~10% of the applied strain, i.e. ~$10^{-3}$)[36]. Beyond this primary creep regime, various recrystallization processes such as dynamic recrystallization start relaxing these internal stresses [12, 16], leading to a stabilization or an increase of the strain-rate under constant stress. These recrystallization mechanisms are out of the scope of the present manuscript.

This kinematic hardening is the signature of the predominance of long-range elastic interactions in the collective dislocation dynamics, and of the concomitant absence of a well-defined dislocation sub-structure with a characteristic scale. We will see below that more detailed analyses fully confirmed these points.

*The effect of impurities/disorder*

In metallurgy, the introduction of impurities (solutes, precipitates,..) has been used for centuries to harden materials[37], as such defects within the crystal lattice have a pinning effect on dislocation motion. In case of diffusing (i.e. mobile) solutes, the interplay between dislocation motion and this slow diffusion of solutes towards dislocations can lead to dynamic strain aging phenomena characterized by a large variety of intermittent regimes of plasticity, which have been thoroughly detailed elsewhere (e.g. [38, 39], and many others). This situation, which is irrelevant for ice, is not considered here.

As shown below, in FCC metals with immobile impurities, their pinning effect frustrates collective dislocation dynamics, hence reduces the intermittency of plastic deformation[40]. Ice is, once again, paradoxical in this respect. First, protonic disorder may actually favor dislocation mobility [29]. Then, the solubility of foreign species within ice Ih lattice is very low, meaning that the solute concentration is always extremely small, with the exception of ammonium fluoride $NH_4F$ [16]. However, even tiny solute fractions (few p.p.m) of HF, HCl or $H_2SO_4$ have a strong *softening* impact on ice plastic flow at the global (crystal, sample) scale [41, 42]. On the other hand, at least for HF, such impurities do not have a significant effect on the mobility of individual dislocations[43]. This suggests that the effect of solutes at the global scale might be through an increase of dislocation density [42], maybe from the activation of additional dislocation sources. We should also note that precipitates do not exist in ice.

In conclusion, compared with FCC or BCC metals or alloys, ice plasticity appears as particularly exotic, characterized by a very strong plastic anisotropy, a very low lattice friction, kinematic hardening instead of isotropic, which likely signs the absence of development of a well-defined dislocation sub-structure and of an emergent internal scale (see more below), and disorder-softening instead of hardening. As discussed below, these differences with classical metallurgical materials are reflected in the nature of plastic "flow", and particularly its intermittency and spatial heterogeneity.

## 3. Intermittency of plastic deformation

### 3.1 Plastic intermittency and dislocation avalanches: historical studies

Historically, the first evidence of jerky, intermittent plasticity was reported in 1932 by Orowan and Becker on Zinc single crystal rods[44]. These authors reported strain curves consisting in a succession of sudden strain jumps, which they already designated as dislocation *avalanches*, i.e. as fast and coordinated motions of numerous dislocations (Fig. 1a). Interestingly, this seminal work likely played an important role in the elaboration of the classical dislocation theory developed by Orowan few years later [45-47]. Later on, torsion tests on tubular rods of monocrystalline Zn confirmed such jerky dynamics[48]. From their highly resolved measurements, these authors revealed that purely elastic loading ramps separated the plastic jumps, suggesting that dislocation avalanches accounted for most of plastic deformation in these samples. Hence, from these early works, signs of plastic intermittency were essentially observed in hexagonal close packed (HCP) materials characterized, like ice Ih, by a strong plastic anisotropy.

As any sudden local change of inelastic strain[49], dislocation avalanches generate (micro)seismic waves that are called Acoustic Emissions (AE) [50, 51]. AE can therefore be used as an efficient probe for such plastic transients. The quantitative interpretation of AE measurements in terms of plastic activity will be given in more details below. Here we simply mention that from the early 70's several authors reported AE monitoring of plastic deformation. James and Carpenter [52] observed well defined acoustic bursts during the monotonic compression of LiF (cubic), NaCl (FCC) and Zn (HCP) single crystals, with a maximum burst rate at macroscopic plastic yield, followed by a rapid decay of AE activity. They attributed these bursts to dislocation "breakaways" that we can now identify as avalanches. This work suggested that (i) AE can indeed be an efficient monitoring tool of plastic intermittency and (ii) dislocation avalanches might not be restricted to hexagonal materials (note, however, that LiF and NaCl strongly differ from FCC metals in terms dislocation core structure, lattice friction, ect..). Other authors also reported some acoustic bursts during plastic deformation of FCC metals such as Copper[53] or Aluminum[50] (Fig. 1b). However, these bursts were very limited in number, occurred essentially near plastic yield, and then almost disappeared in the post-yield, strain-hardening regime. In these materials, most of plasticity-related AE energy is released through the so-called continuous AE consisting in a slow evolution of the background AE "noise", without well-defined transient bursts or "quakes". The interpretation of this continuous AE is that plastic deformation results from the cumulative effect of numerous but *uncorrelated* (in space as well as in time), *independent*, and *small* dislocation motions[51, 54]. Hence, in terms of collective dynamics, this fundamentally differs from dislocation avalanches. It has been recently demonstrated that such continuous AE, which is generally maximum at plastic yield and slowly decays afterwards, has indeed all the characteristics of Gaussian uncorrelated noise, without any sign of intermittency[17].

Taken altogether, these seminal works suggested that plastic deformation can take place through dislocation avalanches, apparently in a rather systematic way in hexagonal materials where these coordinated motions may account for a large portion of total plasticity. On the other hand, in FCC metals, such plastic bursts are much more episodic and disappear during strain-hardening, while the global dynamics seems dominated by the cumulative effect of small uncorrelated motions. In these initial studies, however, a detailed characterization of plastic intermittency in terms of time correlations, distribution of avalanches sizes, or spatial localization, was not performed. After these seminal analyses, this topic was nearly abandoned for few decades. This can likely be due to the concomitant development of classical plasticity theory that more or less implicitly assumes, inspired by smooth stress-strain curves at the macro-scale, that such plastic fluctuations average out at some large enough time and spatial scales (hence defining a representative volume element for plasticity). By the end of the last century, new AE measurements in ice showed that such assumption is wrong, at least in this material, and revived this important subject[55-57].

3.2    Intermittent plasticity in ice

The first evidences of AE generated by dislocation avalanches in ice single crystals were reported more than 20 years ago[55]. The plastic origin of this AE can be ascertain with great confidence. Indeed, as twinning does not occur in ice, the only possible sources of AE are (i) dislocation motion and (ii) crack nucleation and/or propagation. Owing to the perfect transparency of ice single crystals (Fig. 2a), the presence of cracks can be easily detected, making the plastic origin of AE in crack-free specimen unambiguous. A strong correlation between the measured average strain-rate and the average AE activity rate confirmed this plastic origin[55]. Later on, such experiments were repeated under various loading modes (compression or torsion creep, constant stress-rate compression), applied stresses, or temperatures[33, 56, 57]. In all these cases, the intermittent AE signal consisted of a succession of well-defined AE bursts with various amplitudes (Fig. 2b). Before to detail the characteristics of this AE activity in terms of dissipated energy, space and time correlations, the AE source models used to interpret the AE signal in terms of dislocation motion and plasticity should be briefly recalled. More details can be found elsewhere [17, 33].

*AE source models*

The waveform of a typical AE burst recorded during the plastic deformation of ice is shown on Fig. 2c. AE bursts are automatically detected when the signal $A(t)$ overcomes a fixed amplitude threshold $A_{th}$, defining the arrival time $t_0$ of the event, $A(t_0) = A_{th}$. The most commonly used AE characteristics of an individual waveform are the maximum amplitude $A_{max}$ (in absolute value), and the AE energy, calculated as the integrated squared amplitude over the event duration $T$, $E = \int_T A^2(t)\,dt$. The duration is given by $T = t_e - t_0$, where the ending time $t_e$ is obtained when the signal remains below $A_{th}$ over a timescale larger than a pre-defined hit lockout time (HLT), see Fig. 2c.

Considering the case of AE sensors responding to surface velocity, Rouby et al. [51] showed that the acoustic wave amplitude can be related to the cumulated dislocation length $l_d$ of the $n$ dislocations involved in the dislocation avalanche and to their average velocity $v$, $A(t) \sim b l_d v(t)$, where $b$ is the Burgers vector. This quantity accounts for the area $S$ swept in a unit of time by the dislocations during the avalanche, i.e. $A(t) \sim b l_d v(t) = b\frac{dS}{dt}$. Once normalized by the sample volume, this represents a plastic strain rate. To go further and estimate the incremental strain induced by the avalanche, an additional hypothesis on the evolution of $v$ during the avalanche is required. Richeton et al. [33] assumed an exponential decay, $v(t) = v_0 exp(-\alpha(t - t_0))$, as the result e.g. of phonon drag [58], where $\alpha$ is a damping coefficient. Considering a short rise time, i.e. $A_0 \approx A_{max}$ (see Fig. 2c), and integrating this evolution over time, one obtains an AE source model that directly relates the maximum amplitude $A_{max}$ to the total surface swept out by the dislocations during the avalanche, and so, when normalized by the sample volume $V$, to an incremental plastic strain $\varepsilon_p$, $A_{max} \sim bS/V \sim \varepsilon_p$. This exponential decay assumption was supported by a good correlation between the AE activity (defined as $A_{max}$ cumulated over the AE events population) and the global strain during creep tests on ice single crystals [55], as well as by a $T \sim ln(A_{max})$ scaling [33]. However, the last relation might also result from the scattering of acoustic waves throughout the material and their reflection at the sample surfaces, generating an exponentially decaying coda (e.g. [59]). This stress the fact that (i) the interpretation of the AE event duration $T$ as the genuine dislocation avalanche duration and (ii) the proportionality between $A_{max}$ and the incremental strain $\varepsilon_p$, have to be taken with caution. Whatever it be, $A_{max}$ remains a useful proxy of the "size" of the dislocation avalanche.

However, in other systems exhibiting intermittent slip avalanches, such as sheared granular media[60] or faults[61], the radiated acoustic/seismic energy $E$ is considered as being proportional to the dissipated energy at the source, hence to well characterize the corresponding avalanche. In order to allow a better comparison with those systems, we will focus here mainly on AE energies instead of amplitudes. As $E \sim A_{max}^2$ [62], these two descriptions of the avalanches (either from $E$ or $A_{max}$) are equivalent. Note that this $E \sim A_{max}^2$ scaling is compatible with the exponential decay hypothesis mentioned above, but with other assumptions as well, such as a constant velocity $v$ during the avalanche.

When plastic deformation does not take place through intermittent, well-defined dislocation avalanches, but instead results from the cumulative effect of numerous uncorrelated and small dislocation motions, continuous AE is generated [51, 54]. Considering that these independent but similar AE sources are characterized by a dislocation sweeping area of mean value $\langle s \rangle$, the resulting AE power scales as $\frac{dE}{dt} \sim \frac{dN}{dt} b^2 \langle s \rangle^2 \sim b \langle s \rangle \frac{d\varepsilon_p}{dt}$, where $\frac{dN}{dt}$ is the number of sources activated per unit time, i.e. the AE power is a measure of the plastic strain-rate[63, 64]. As shown below, such continuous AE is negligible in ice (see also Fig. 2b), but dominates the AE signal in FCC metals[17].

*Dislocation avalanche energy distributions*

Figure 3a shows the probability density function (PDF) of the acoustic burst energies recorded during the compression creep of an ice single crystal under various applied stress $\sigma$ [57]. The remarkable feature on this figure is the power law distribution of energies,

$$P(E) \sim E^{-\kappa_E} \quad (1)$$

, over about 7 orders of magnitude, without detectable upper cut-off, while the lower cut-off results from the experimental detection threshold. For uniaxial compression experiments, the exponent $\kappa_E$ was found to be independent of the applied stress (Fig. 3a), of temperature [33], or of the orientation of the c-axis vs the loading axis, with $\kappa_E = 1.40 \pm 0.03$. The exponents were estimated from a robust maximum likelihood methodology [65], and the associated uncertainty accounts for the variability between different tests, applied stresses, temperatures, or c-axis orientation[17]. As $E \sim A_{max}^2$, this power law distribution of energies translates into a distribution of amplitudes, $P(A_{max}) \sim A_{max}^{-\kappa_A}$, with $\kappa_A = 2\kappa_E - 1 = 1.80$. Such distributions have several fundamental consequences:

-        It implies that plastic "flow" in ice is scale invariant, with an absence of characteristic scale over at least 7 orders of magnitude for dissipated energy, or 3 ½ orders of magnitude in slip "size".

-        As $\kappa_E$ and $\kappa_A$ are smaller than 2, both the mean and the standard deviation of $E$ and $A_{max}$ are theoretically undefined. In more practical terms, this means that most of the energy and strain are dissipated through few, large avalanches, while the remaining population of plastic fluctuations accounts for a negligible part of total deformation.

- A combination of the two points mentioned above precludes the definition of a representative volume element (RVE) for ice single crystal plasticity.

- These distributions are the signature of a dynamical system at (or near) a critical state. We will re-discuss below the possible nature of this criticality, which is still debated nowadays[66, 67]. However, as the distribution is independent of the applied stress and the stage of deformation, it can hardly be associated to a critical point reached through the fine tuning of the control parameter (the applied stress).

*Time correlations*

These dislocation avalanches do not occur randomly through time, instead are characterized by time-clustering, another clear signature of the intermittency of ice plasticity[62]. More precisely, the avalanche rate, $dN/dt$, immediately after any avalanche is, in average, larger than the background activity due to uncorrelated events[68] (Fig. 4). This self-induced triggering increases, and remains above the background rate for longer times, as the energy of the mainshock increases. Such aftershock triggering results from stress redistributions following plastic avalanches, and surprisingly mimics crustal seismicity[69]. This process is asymmetric in time, as no significant foreshock activity can be detected before avalanches. Hence, the avalanche rate can be written as:

$\frac{dN}{dt} = B(t) + \sum_{i:t_i<t} f_T(t)$ (2)

, where $B(t)$ is a background rate accounting for uncorrelated events (a Poisson process), $f_T(t)$ is a function describing the aftershock triggering rate, while summation is restricted to all events occurring at times $t < t_i$. Note that the interpretation of the continuous AE mentioned above assumes the absence of any related triggering, i.e. $dN/dt$ reduces to a background activity in this case.

For crustal seismicity, it has long been recognized that the aftershock triggering rate decays as $f_T(t) \sim \frac{1}{(t+c)^p}$ where $p$ is generally close to 1 and $c$ is a (small) time constant; this is the so-called Omori's law[69, 70]. Such identification is favored in this case by low background rates. However, in the creep experiments performed on cm-size ice single crystals (Fig. 2a), $B(t)$ is large (about 170 events/s on Fig. 4), hence the determination of $f_T$ is difficult. However, it has been shown that larger avalanches trigger, in average, more aftershocks than smaller ones, following the productivity law $N_M \sim A_{max}{}^\alpha \sim E^{\alpha/2}$, where $N_M$ is the number of aftershocks triggered per mainshock and $\alpha = 0.6$ [68]. This productivity law is reminiscent, once again, of what is observed for earthquakes [71].

*Spatial organization*

The power law distribution of avalanche energies as well as the aftershock triggering process described above are clear signatures of a collective dislocation dynamics governed by long-ranged elastic interactions between dislocations [57]. Such dynamics is also characterized by a spatial clustering of the avalanches following a fractal pattern (Fig. 5) [72]. This spatial clustering is linked to the above-mentioned time clustering through a space/time coupling: the closer in time two avalanches are, the larger the probability is that they will be closer in space. This is another signature of a triggering process resulting from internal stress redistributions following plastic events. As such stress redistributions decay with the distance from the mainshock, triggering is naturally associated with spatial clustering, while the space/time coupling results from a cascade process in which avalanches positively modulate the occurrence probability of the subsequent avalanches in their vicinity, which in turn modify the occurrence probability of the third generation of avalanches, and so on, at increasing distances (on average) from the avalanches of the first generation. Such "diffusion" of the activity (Fig. 5c) and space/time coupling have been also documented for earthquakes [71, 73-75], raising again a surprising analogy between ice plasticity and crustal deformation.

This fractal patterning reveals the absence of a spatial characteristic scale in the plastic deformation of ice single crystals, much like the power law distribution of energies (Eq. (1)) illustrates the absence of a specific avalanche "size". We show below that the presence of grain boundaries (GBs) partly modifies this assessment.

These observations are fully consistent with the analysis, from synchrotron topography, of slip lines and dislocation arrangements in ice single crystals deformed under torsion[25]. In X-ray topography, the diffracted intensity records the lattice distorsion, related in this case to the dislocation density. Profile analysis of the topographs revealed scale-invariant dislocation density patterns with long-range correlations, expressed by a power spectrum of the profiles scaling as $p(f) \sim f^{-\mu}$, with μ=1.3±0.1, over more than two orders of magnitude (from ~15 μm to ~7 mm), i.e. up to almost the sample size scale. Unlike AE measurements, these observations are "static" in nature, revealing the spatial correlation of dislocations "at rest". The emergence of such scale-invariant patterns, which were also observed in Cd single crystal, a HCP metal[76], can be explained by long-ranged elastic interactions between screw dislocations and multiplication through cross slip on prismatic planes[26].

*Effect of temperature*

Richeton et al.[33] analyzed the role of temperature on dislocation avalanches from AE measurements in ice single crystals. In the temperature range explored (-20°C≤ $T$ ≤-3°C), relevant for most laboratory and terrestrial conditions, the power law distributions of avalanche amplitude or energy as well as the time clustering of avalanches (see above) were found to be temperature independent. In terms of homologous temperature $T_h = T/T_m$, these temperatures remain large ($T_h > 0.92$). However, a comparison between ice and e.g. classical metallic materials in terms of $T_h$ would be misleading as e.g. diffusional creep (Coble or Nabarro-like) is irrelevant in ice, even close to the melting point [29].

A well-known effect of temperature is on the velocity of an *individual* dislocation under a *low-velocity* regime (~μm/s), which follows an Arrhenius scaling, $v_i \sim \sigma exp(-Q/kT)$, with $Q \approx 0.9$ eV [28, 29]. However, this low velocity regime is most liklely irrelevant in case of dislocation avalanches [33]. Indeed, such velocities would not generate inertial effects, and so AE. In the case of very fast moving dislocations (generating AE), the characteristic time scale involved in thermally activated processes is likely too large for such processes to be efficient. Instead, in this regime, the predominant temperature-dependent resistance to dislocation motion is no more the overcoming of local obstacles but the drag resistance resulting from the interaction between moving dislocations and excitations, such as phonons [58, 77]. Phonon drag resistance is known to increase with temperature [58, 77]. Therefore, in a high velocity regime, the dislocation velocity is expected to scale with temperature in the opposite way as it does in a thermally activated region. Richeton et al. [33] showed that, for a same AE amplitude, i.e. a same initial plastic "strain-rate" (see above), the average duration of an avalanche is smaller at higher *T*, or, in other words, the damping/drag coefficient $\alpha$ is larger. This observation is consistent with the phonon drag scenario. However, this drag coefficient does not have a significant effect on dislocation avalanches and plastic intermittency. Indeed, collective dislocation dynamics in ice is, as mentioned above, essentially ruled by long-ranged elastic interactions, which nature is not supposed to change with temperature.

*Polycrystals*

All the observations reported above were obtained for ice single crystals. The presence of grain boundaries (GBs) is expected to play a significant role on this collective dynamics. Indeed, GBs act as strong barriers to dislocation motion, as evidenced e.g. by dislocation pile-ups [78], leading to strong internal stresses and kinematic hardening in ice polycrystals [6, 34]. We thus expect GBs to frustrate the propagation of dislocation avalanches beyond the grain size. On the other hand, these internal stresses may activate nearby dislocation sources in neighboring grains[79], or the GBs themselves can act as dislocation sources[80].

AE recording during compression creep of ice polycrystals with an isotropic texture and an average grain size $\langle d \rangle$ varying from 260 μm to 5 mm revealed that GBs indeed hinder the propagation of dislocation avalanches [81] (Fig. 6a). This is expressed by an upper cut-off in the power law distribution of AE burst amplitudes, $A_c$, which decreases with decreasing grain size following $A_c \sim \langle d \rangle^{2.4 \pm 0.3}$. This non-trivial scaling might result from an anisotropic, "lamellar" internal structure of the dislocation avalanches characterized by a fractal dimension significantly smaller than 3 [82], consistently with the anisotropic character of ice plasticity. Whatever it be, these truncated power law distributions (Fig. 6) indicate that avalanches are dynamically confined within individual grains, with possible important consequences. Indeed, compared with other materials, grain sizes are generally very large in ice. Hence, in materials with grain sizes in the μm range, such mechanism would make plastic intermittency almost undetectable, and a volume encompassing several grains might be considered as representative (RVE) of the material plasticity. However, the observations performed on ice polycrystals suggest a more complex scenario. Besides an upper truncation, the presence of GBs lowers the power law exponents of avalanche size distributions, $\kappa_A$ and $\kappa_E$. In addition, much like in single crystals (see above), avalanches are spatially organized following a fractal pattern up to a length scale at least one order of magnitude larger than $\langle d \rangle$, and possibly only limited by the system-size [33] (Fig. 6b). These observations can be rationalized as follows: the confinement of dislocation avalanches within grains builds up internal stresses that push temporally the system into a supercritical state, off the scale invariant regime of single crystals, and trigger secondary avalanches (aftershocks) in neighboring grains[81]. In other words, if GBs hinder the dynamical propagation of individual avalanches, correlated plastic activity spreads over much larger distances. Such scenario seems consistent with recent molecular dynamics simulations of plasticity using simplified polycrystal models [83].

The connection between dislocation avalanches and kinematic hardening was also explored upon unloading polycrystalline samples [33]. During the plastic strain recovery characterizing this unloading (see section 2), a relatively brief (few s) but strong AE activity was recorded. This can be interpreted as dislocations starting gliding back from pile-ups at boundaries in an avalanche-like manner. This is consistent with a power law size distribution of these avalanches without upper cut-off, and with an exponent consistent with size distributions in single crystals [33]. In other words, during this strain recovery, dislocations gliding back from pile-ups "do not feel" the opposite boundary.

## 3.3    Comparison with other materials and the definition of wildness

The observations summarized above show that plastic deformation in ice occurs in a strongly intermittent way, through dislocation avalanches clustered in both time and space and power law distributed in size (incremental strain) and energy. The values of the corresponding exponents ($\kappa_A$ and $\kappa_E$ <2) imply that most of plastic deformation takes place through few, very large avalanches that control the global dynamics. Following a terminology first introduced by B. Mandelbrot for financial fluctuations[84], one can identify such dynamics as *wild*, contrasting with a *mild* dynamics that would be characterized by small, essentially uncorrelated fluctuations  which can be homogenized at some mesoscopic time- and spatial-scales. In more quantitative terms, the *wildness* of plasticity, $W$, can be defined as the fraction of deformation dissipated through power-law distributed dislocation avalanches[17]. Using AE to probe plastic flow dynamics and associated fluctuations, the ratio between the AE energy cumulated over all the detected bursts in one hand, and the total AE energy recorded over the entire deformation process on the other hand (once the instrumental noise removed), represents a proxy of this wildness, $W_{AE}$. In ice, $W_{AE}$ was found to be always larger than 0.999 whatever the loading mode, the temperature, or the texture (single- vs poly-crystals)[17]. This independence of wildness vs temperature, at least in the range explored by experiments (-20°C$\leq T \leq$-3°C), is fully consistent with what has been said in section 3.2. This shows that ice can be considered as the paradigm of wild plasticity.

A comparison with other materials and crystalline structures is enlightening to unravel the origin of this wildness. Similar AE monitoring of plastic deformation was performed during monotonic tensile tests on HCP (Zn-0.08%Al and Cd) and FCC (Cu, Al and Cu-Al) metals and alloys[17]. HCP materials exhibit an intermittent plastic flow with power law distributed avalanche sizes and energies, and wildness $W_{AE}$ always larger than 0.95. This close similarity with ice points out the role of slip anisotropy at the individual dislocation scale to promote wildness at the global scale.

This fully contrasts with the behavior of FCC materials. In this case, although the episodically recorded AE bursts are power distributed as well, the global activity is dominated by a continuous AE signal, signature of mild, uncorrelated fluctuations. Consequently, $W_{AE}$ is in the range 0.1-0.5 for pure Copper, 0.01 to 0.1 for Cu-Al alloys (a first indication of the role of extrinsic disorder – solutes in this case – on wildness), and 0.01 to 0.02 for pure Al [17]. In addition, cyclic loading tests on Al, a loading mode known to promote well-defined dislocation sub-structures such as cells and walls in association with cyclic strain hardening[85], revealed a rapid decrease of wildness with the number of cycles, and a quasi-suppression of any detectable avalanche after few hundreds of cycles[86]. These results demonstrate that single-slip plasticity, ruled by elastic long-range interactions between dislocations and kinematic hardening, promotes wildness, while in multi-slip systems such as FCC metals, plasticity is essentially governed by short-range/forest interactions leading to the emergence of dislocation sub-structures, isotropic hardening, and mildness. In classical plasticity theory, these sub-structures are considered as dynamically stable. However, in the experiments mentioned above for FCC materials, mild and (few) wild plastic fluctuations do *co-exist*, with the latter being associated with rare but extended and brutal rearrangements of the dislocation sub-structure [17, 87]. In these materials, these wild fluctuations are still power law distributed, though with larger exponents $\kappa_A$ and $\kappa_E$ compared with ice [17]. As shown below, this association larger mildness/larger exponents is not incidental, but the signature of an underlying "universal" framework. It is worth noting that, when mild and wild fluctuations co-exist, the power law distributions of avalanche sizes are *lower*-truncated by the presence of a characteristic size at small scales. This fundamentally differs from the upper-truncations induced by the presence of GBs (see above and Fig. 6a).

AE is an indirect probe for plastic fluctuations. As mentioned in section 3.1, the first evidences of dislocation avalanches were obtained directly from the stress-strain curves of thin rods[44]. At this stage, a question arises: Why do we not detect similar sudden strain steps during mechanical tests on ice crystals ? The answer is likely twofold. It might be first related to a too low sampling frequency and sensitivity of the strain sensor (~10 μm of resolution over a stroke of 5 mm for classical LVDT sensors, i.e. a ~$10^{-3}$ strain sensitivity). Sampled at 1 Hz, the sensor will probe the cumulated effect of hundreds of avalanches, which, for most of them, belong to the background activity $B(t)$, hence are not concerned by the space/time coupling of plastic activity mentioned in section 3.2. In addition, as suggested by the non-trivial scaling between the cut-off amplitude $A_c$ and the grain size in polycrystals (see section 3.2) and in agreement with the presence of strong slip localization [25], plastic avalanches in ice are likely not dense and isotropic, but instead characterized by a fractal structure of dimension $D < 3$. In this case, finite size effects (the avalanche cannot extend more than the sample size $L$) imply that the maximum volume strained by an avalanche should scales as $L^D$, hence the corresponding strain increment as $L^{D-3}$ [82]. Therefore, this cut-off strain increment decreases with increasing sample size, explaining the smoothness of macroscopic deformation curves.

The same reasoning implies that wild plastic fluctuations should become unescapable in very small structures (few μm and below). Such mechanical experiments were not (yet) performed on ice. However, with the recent development of nanotechnologies, the physics of plasticity at such small system sizes became a subject of growing interest over the last 15 years [88, 89], and the micropillar compression test a standard[90, 91]. Loading curves with sudden stress drops (strain-controlled tests) or strain jumps (stress-controlled tests), corresponding to strain avalanches, were reported in various materials such as HCP materials [92] and alloys[93], FCC metals[94] or alloys[40], or BCC metals [95, 96]. The first statistical analyses of the incremental strains associated with these stress drops or strain jumps, which can be directly derived from the stress-strain curves in this case, argued for a power law size distribution in pure metals [94, 96, 97],

$$P(\varepsilon) \sim \varepsilon^{-\kappa_\varepsilon} \ (3)$$

, with $\kappa_\varepsilon \approx 1.6$, i.e. rather close to the exponents reported for AE amplitudes or energies in ice[94]. These observations pointed out an unexpected result: Materials exhibiting a mild plastic flow at bulk scales (see above) are characterized by an intermittent deformation at μm- to nm-scales, revealing a size effect on wildness[40]. Qualitatively, micropillar stress-strain curves appear less jerky upon increasing the sample size above few μm [95].

A recent study analyzed more systematically this size effect on wildness, for Al and Al-alloys[40]. In this case, the true wildness $W$ can be measured from a detailed analysis of the stress-strain curves. The main conclusions of this work can be summarized as follows:

(i) The wildness $W$ decreases with increasing sample size, from $W \approx 1$ for very small systems (below 500 nm for pure Al) to $W \approx 0$ above few μm. Hence, these experiments allow to cross over the entire wild-to-mild transition.

(ii) An increasing wildness is associated with an increasing slip anisotropy. In other words, in these FCC materials, plasticity switches from multi-slip to single slip upon decreasing the sample size.

(iii) Consequently, Taylor (forest) hardening is suppressed at small scales[98].

(iv) The exponent $\kappa_\varepsilon$ is not universal, instead is tightly linked to the wildness $W$ through a relationship that itself appears material-independent: the smaller the wildness, the larger the exponent (Fig. 7). Note that, besides this relation, a larger exponent means that the largest avalanches become so rare that they accounts for a less and less significant part of global plastic strain.

(v) The introduction of an internal disorder in the crystalline structure through alloying (solutes or precipitates) counteracts the external size effect: It promotes a mild plasticity, slip isotropy, and forest hardening.

From this comparison between different materials, at various system sizes, the wildness of ice plasticity, and of other materials for small enough external sizes, can now be unambiguously ascribed to slip anisotropy, which promotes the role of long-ranged elastic interactions in collective dislocation dynamics while suppressing isotropic hardening related to short-range interactions. In FCC materials, such slip anisotropy, hence wildness, is only encountered at µm- and sub µm-scales, when surface effects and dislocation starvation mechanisms [99] take place. Owing to the crystalline structure, in HCP materials, and particularly in ice Ih, slip anisotropy, an absence of forest hardening, and wildness remain at bulk scales. In other words, one may say that large ice single crystals (>> mm) can be considered, in terms of the physics of plasticity, as "giant micropillars" – an oxymoron.

4. Discussion and conclusions

*Criticality*

The observations reported in section 3 on ice single crystals revealed a surprising similarity between the collective dislocation dynamics in this material, and brittle crustal deformation and earthquakes. Indeed, both systems share a power-law distribution of "seismic" energies and associated slips (the Gutenberg-Richter law in case of earthquakes[100]), time correlations and aftershock triggering quantified by a productivity law, a fractal spatial organization of events, and a non-trivial diffusion of activity resulting from cascades of aftershock triggering. These features, and particularly the absence of characteristic scales, indicate that both systems are in (or close to) a critical state, in which any arbitrarily small and local (stress) perturbation may potentially have a large impact with an amplitude only constrained by the system size itself. In case of earthquakes, the nature of the criticality has been a subject of strong debate for almost 30 years (e.g. [71, 101-103]). In particular, it was proposed that earthquake occurrence could be interpreted as a self-organized critical (SOC) phenomenon[101, 102], corresponding to a system spontaneously evolving towards a critical state, without the necessity of a fine tuning of the control parameter (deformation rate in this case), as long as this rate of elastic energy injection is very slow compared with the rate of energy relaxation during avalanches. On the other hand, the nucleation of individual earthquakes has been mapped onto the problem of the depinning transition of an elastic manifold, hence implying a *tuned*-criticality (scale invariance occurs only at the critical point, i.e. the "failure" stress)[104, 105].

Owing to the strong analogy between the two systems, a similar debate also happened in case of dislocation avalanches. It is worth recalling that the microscale physics of deformation is different in the two systems, as dislocation motion and plasticity is a friction-free process, i.e., unlike earthquake nucleation and propagation, is insensitive to the normal stress. This suggests that the exact nature of the local threshold dynamics is unimportant to define the collective behavior. Instead, the two systems share some ingredients essential for the emergence of avalanche dynamics and intermittency:

- long-ranged elastic stress redistribution following a dislocation avalanche or an earthquake. Note that the associated elastic redistribution kernel, which is non-convex, is similar, as the fault slip associated with an earthquake can be modelled from the continuum theory of dislocations[106].

- presence of disorder

- a slow driving condition with an external driving rate (applied strain-rate on experiments; tectonic loading for earthquakes) much smaller than the rate of avalanche spreading and of elastic stress redistribution.

The AE data obtained on ice single crystals argued for a SOC-like interpretation of plastic deformation, at least in this material and maybe on a more general perspective [57, 107]. However, it has been argued more recently that the dislocation avalanches recorded during the plastic deformation of metallic (FCC and BCC) micropillars might be associated instead with tuned-criticality, i.e. with true scale invariance only reached at the system-size plastic yield $\tau_y$ (the critical point). In this case, upper-truncated power law distributions of avalanche sizes would be observed for applied stresses below $\tau_y$, this truncation signing a finite correlation length much smaller than the system size. This interpretation, however, seems irreconcilable with the absence of detectable upper cut-off in the distributions of AE burst energies in ice crystals, and with the independence of these distributions with the applied stress (Fig. 3). This tuned-critical scenario has also been challenged by 2D and 3D discrete dislocation dynamics (DDD) simulations that argued for a system-spanning correlation length at any applied stress, but an average avalanche size exponentially growing with this stress [67, 108]. It has also been proposed that plasticity could be interpreted as a critical yielding transition, but with a critical point (yielding point) constantly moving as the material strain hardens, thus explaining how intermittent dynamics and critical features are observed over a large range of stress[109]. This scenario, however, cannot explain dislocation avalanche dynamics in ice single crystals, which do not strain harden.

This discussion illustrates the fact that, much like for earthquakes [71], the nature of the criticality of plastic deformation remains a lively question.

*An interpretation of the extreme wildness of ice plasticity*

Beyond this debate, we have seen in section 3 that the intermittency and wildness of ice plasticity appears as an extreme case (see Fig. 7), as the crystal structure (e.g. FCC instead of hexagonal), the system size, the introduction of extrinsic disorder (solutes, precipitates), or GBs may fundamentally alter the collective dynamics.

It was mentioned that multi-slip plasticity, allowing the emergence of dislocation sub-structures, i.e. of an intrinsic disorder, as well as large system sizes and extrinsic disorder, all favor mild fluctuations that are incompatible with a pure critical interpretation of plasticity. In other word, SOC-like criticality of plasticity might be actually restricted to ice, and maybe to other HCP materials. In this context, the relation between wildness and the power law exponent (Fig. 7) might instead illustrate an underlying "universal" framework for plastic fluctuations and dynamics, with the SOC-like behavior of ice being just an end-member. It has been recently proposed that the key parameter controlling the nature of plastic fluctuations in crystalline materials (mild vs wild) is a ratio of length scales,

$$R = \frac{L}{l} \quad (4)$$

, where $L$ is the system size and $l$ an internal length scale[17, 40]. These authors defined $l$ as

$$l = \frac{Gb}{\tau_{pin}} \quad (5)$$

, where $\tau_{pin}$ is the pinning strength of all kind of obstacles to dislocation motion (except GBs; see below), including lattice friction ($\tau_l$) , forest dislocations ($\tau_f$), solutes ($\tau_{sol}$), or precipitates ($\tau_{pp}$). Hence, $\tau_{pin}$ reads:

$$\tau_{pin} = \tau_l + \tau_f + \tau_{sol} + \tau_{pp} \quad (6)$$

In this mean-field description of obstacles, $l$ represents the length scale at which the dislocation-dislocation elastic interaction stress ($\sim Gb$) becomes equal to the dislocation-obstacle interaction stress $\tau_{pin}$ [110]. For pure FCC metals with a low lattice resistance, the effect of forest dislocations, standing on different slip systems, dominates the pinning strength. Hence, in this case, $\tau_{pin} \sim Gb\sqrt{\rho_f}$, where $\rho_f$ is the forest dislocation density, and $R \sim L/l_f$, with $l_f = 1/\sqrt{\rho_f}$ being proportional to the mean free path of mobile dislocations. According to the similitude principle[111, 112], $l_f$ represents the typical length scale of the emerging dislocation sub-structure. In BCC metals, the lattice resistance term $\tau_l$ might play a larger role below the transition temperature, while in alloys $\tau_{pin}$ is essentially set by $\tau_{sol}$ or $\tau_{pp}$ [40]. For FCC micropillars (Al and Al-alloys), the combination of eqs. (4) to (6) allowed a scaling collapse of wildness values for all materials and sample sizes (Fig. 8)[40].

In this framework, the extreme wildness of ice plasticity can now be interpreted as follows. First, extrinsic disorder effects are negligible. In addition, lattice resistance is very small: taking the upper estimate of $10^{-5}$ for $\tau_l/G$ (see section 2) and $b = 0.452$ nm for basal dislocations[16], eq. (5) gives a length scale of ~ 50 μm. In this context, the absence of detectable mild fluctuations in cm-size ice samples actually argue for an even smaller $\tau_l/G$, of the order of $10^{-8}$ or below, as already suspected in section 2. Finally, the plastic anisotropy of ice implies an absence of forest (isotropic) hardening, i.e. $\tau_f$ is negligible as no internal length scale emerges from short-range dislocation interactions (see section 2). Taken altogether, this implies that $\tau_{pin}$ is particularly small in ice, hence $l$ is large (>> mm), elastic interactions rule the dynamics, and plasticity remains wild and critical even at the bulk scale (Fig. 8). This discussion rationalizes the link between plastic anisotropy, low lattice resistance, kinematic (instead of isotropic) hardening, and wildness.

It has been proposed, from a continuum model of plasticity [113] as well discrete dislocation dynamics simulations [114], that strain-hardening induces an upper cut-off in the distribution of dislocation avalanche size. If strong enough, this strain-hardening effect could explain the difference, in terms of dislocation avalanches, between ice single crystals (no strain hardening, hence pure wildness), and bulk FCC samples (strongly truncated distributions of avalanche sizes). Such upper cut-off might also be reminiscent of the above discussion about the role of the length scale $l_f = 1/\sqrt{\rho_f}$ . The physical meaning is, however, different, as the upper cut-off analyzed in [113, 114] is considered to play the role of the grain size in ice (see section 3.2) in limiting the extension of dislocation avalanches. Hence, this approach does not explain the *co-existence* of mild (i.e. Gaussian-like) fluctuations at small scales, and wild (power law distributed) fluctuations at large scales.

*Modeling*

To explain this co-existence of mild and wild fluctuations in plastic deformation and the dependence of wildness on material characteristics or sample size, Weiss et al. proposed a simple mean-field stochastic model of collective dislocation dynamics under constant stress (i.e. consistent with the creep tests on ice samples mentioned above) [17]:

$$\frac{d\rho_m}{d\gamma} = A - C\rho_m + \sqrt{2D}\,\rho_m\xi(\gamma) \quad (7)$$

, where $\rho_m$ is the spatially-averaged mobile dislocation density, $\gamma$ the average shear strain, $A$>0 the net nucleation rate, $C$>0 the rate of mutual annihilation/immobilization of dislocation pairs. In this simplified framework, long-ranged stochastic interactions are described by a multiplicative noise involving a standard white noise $\xi(\gamma)$ with zero average and delta-type correlations. The noise intensity $D$ quantifies the intensity of fluctuations experienced by a representative volume due to interactions with the rest of the system. The stationary probability distribution for $\rho_m$ predicted by this model is characterized by mild fluctuations around a well-defined mode $\rho_m = A/(C + D)$ at small sizes, coexisting with a power law tail with exponent $\kappa = 1 + \frac{C}{D}$ at large sizes, much like the distributions of strain fluctuations detailed in section 3. The relative importance of the power law tail, measured by the wildness $W$, was shown to depend only on $C/D$, thus arguing for a universal $W(\kappa)$ relationship: the larger the $C/D$ dimensionless ratio, the larger the exponent and the smaller the wildness. In this context, the specificities of ice plasticity translate as follows: owing to the absence of forest hardening, the immobilization of dislocation pairs is negligible, hence $C$ is small. In addition, as $\tau_{pin}$ is small for the reasons detailed above, $D$ is large. This combination leads to a small $\kappa$, and a large wildness.

*Grain boundaries*

In the framework discussed above, the effects of GBs were not considered. In other words, the internal scale $l$ was always considered to be smaller than $\langle d \rangle$. Indeed, dislocations do not overcome GBs as they do for Peierls barriers, forest dislocations, solutes or precipitates, and a mean-field description of these GBs effects on dislocation motion and dynamics appears problematic. However, it has been shown in section 3.2 that GBs do hinder the propagation of individual dislocation avalanches, whereas correlated plastic activity spreads over distances much larger than the grain size. Based on this transmission of plastic deformation (but not of dislocations) through GBs revealed on ice polycrystals [81], Louchet et al.[115] proposed a simple model to account for the well-known Hall-Petch law describing the grain-size dependence of the macroscopic plastic yield in polycrystals[116, 117],

$$\tau_y = \tau_\infty + KG(b/d)^{1/2} \quad (8)$$

, where $\tau_\infty$ is the plastic yield of single crystals and $K$ a constant in the range 0.05-0.5. This model is based on the idea that the macroyield $\tau_y$ corresponds to a stress level for which the incompatibility stresses between neighboring grains with different orientations can be relaxed through a system-wide transmission of strain bursts, letting strain to "percolate" through the material. As already mentioned, $\tau_\infty \sim \tau_l$ is very small in ice. Although the grain-size dependence of $\tau_y$ in ice polycrystals from monotonic loading tests is still unexplored,

Duval and co-workers [118, 119] reported a decreasing primary creep rate with decreasing grain size that might be related with such Hall-Petch mechanism [16].

This exploration of plastic intermittency in ice, and the comparison with other materials, revealed the very exotic character of this material in this respect, in close relation with its extreme plastic anisotropy already suspected by the end of the XIXth century and fully characterized almost 40 years ago[6]. For a very long time, the main purpose of ice plasticity studies was a better understanding and modeling of glaciers and ice sheets. The present story exemplifies the relevance of ice as a model material to deepen our understanding of the mechanical behavior of materials in general[120]. The growing interest on dislocation avalanches and the nature of associated fluctuations in the Materials Science literature, triggered by AE measurements performed on ice single crystals 20 years ago, is, I believe, a good illustration.

## Additional Information

**Competing Interests**
The author declares that he has no competing interests.

**Acknowledgments**
The author thanks Paul Duval, whose founding work on ice plasticity was a great source of inspiration, Maurine Montagnat for interesting discussions on the manuscript, and Peng Zhang for the preparation of Fig. 7. Two anonymous referees are also thanked for interesting comments and suggestions.

# Figures

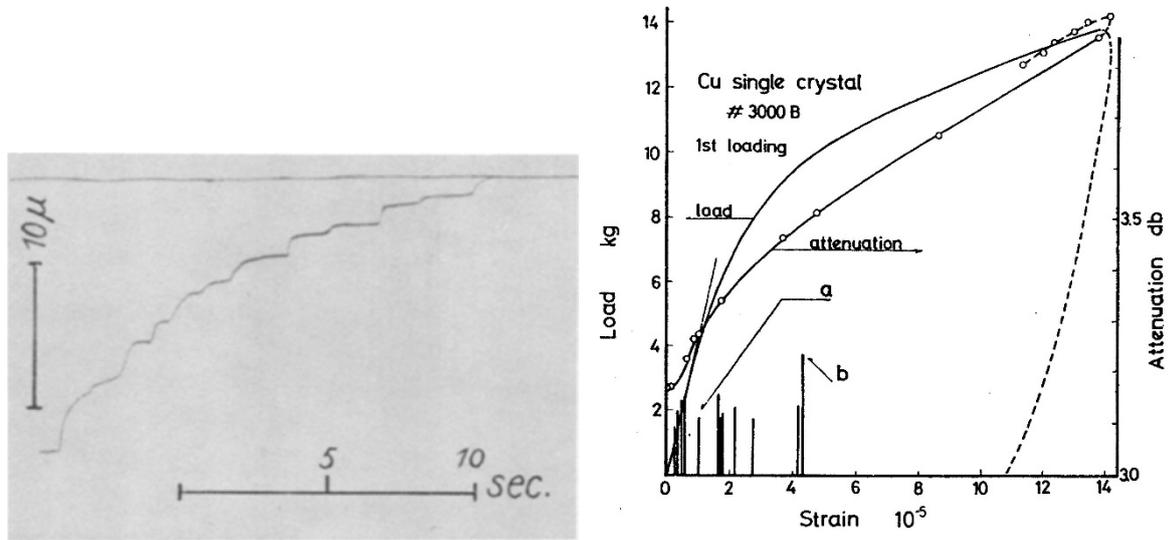

*Fig. 1.* Historical evidences of intermittent plasticity. (a) Intermittent strain jumps during the plastic deformation of a Zn single crystal at 80°C, from [41]. (b) AE bursts during the initial loading phase of a copper single crystal under uniaxial tension ($\dot{\varepsilon} = 2 \times 10^{-5}$ s$^{-1}$). This intermittent activity ceases after the macroscopic plastic yield (from [50]).

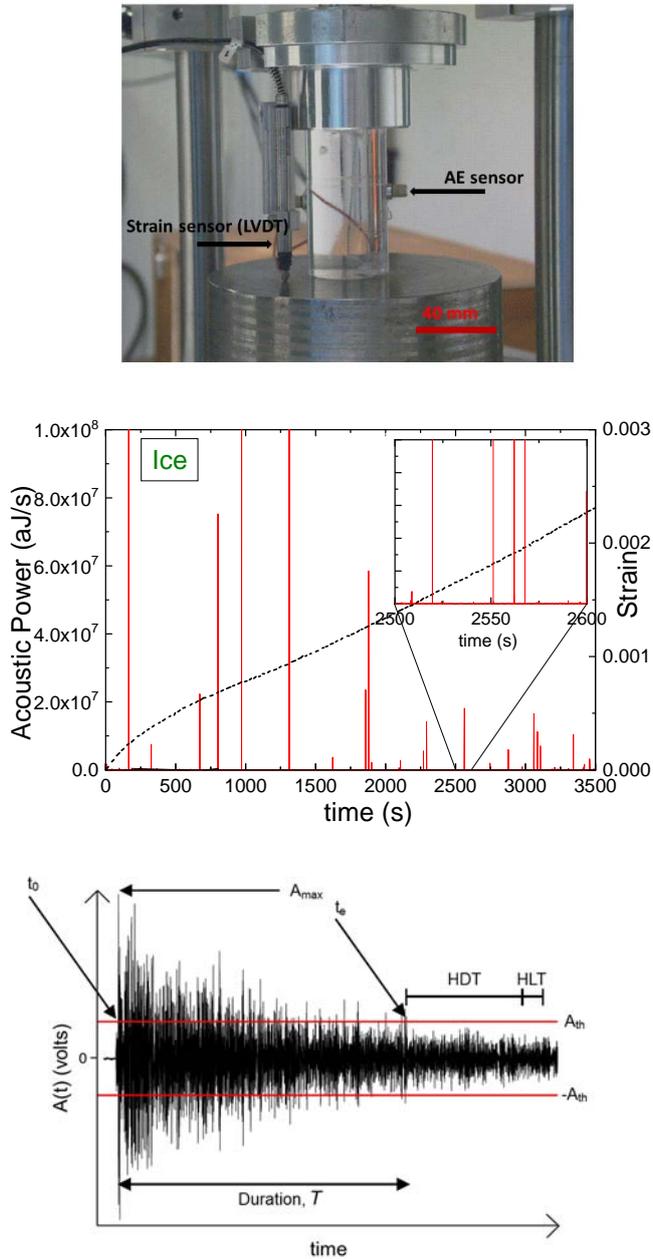

*Fig. 2*. Acoustic emission during viscoplastic deformation of ice single crystals. (a) Experimental set-up for a compression creep test (constant uniaxial stress). (b) Typical intermittent AE signal (Acoustic power sampled at 100 Hz) during a creep test, $\sigma = 0.56$ MPa, Temp.= -10°C (in red). Black dashed line: uniaxial strain. (c) A typical recorded AE waveform corresponding to a dislocation avalanche, with the definition of the main AE characteristics (see text for details).

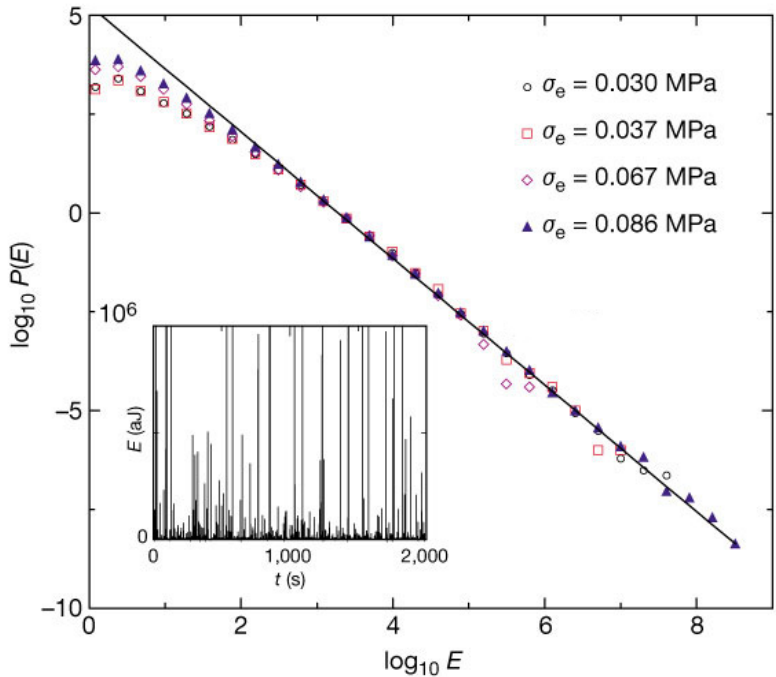

*Fig. 3*. Probability density function (PDF) of the acoustic burst energies recorded during a compression creep test on an ice single crystal at -10°C under various resolved shear stress on the basal plane (from [54]).

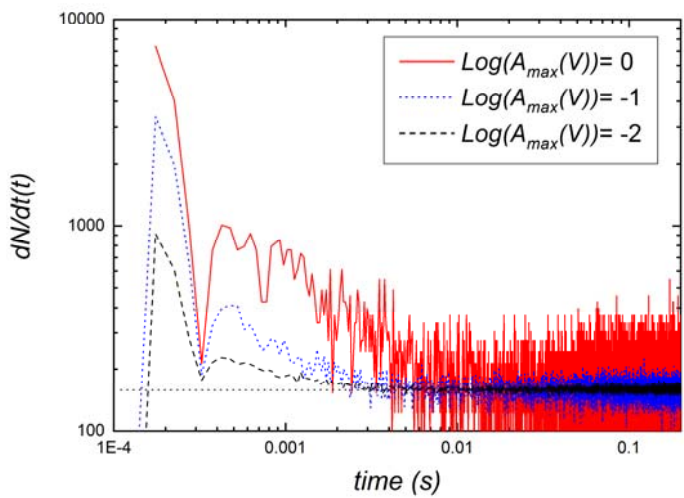

*Fig. 4*. Time correlations of dislocations avalanches: Average AE event rate, $dN/dt$, recorded after dislocation avalanches of "magnitude" $Log(A_{max}) \pm 0.25$, per mainshock and per unit of time, during a compression creep test on an ice single crystal (-10°C; resolved shear stress on the basal plane: 0.086 MPa). The background activity is represented by the horizontal dashed line (adapted from [65]).

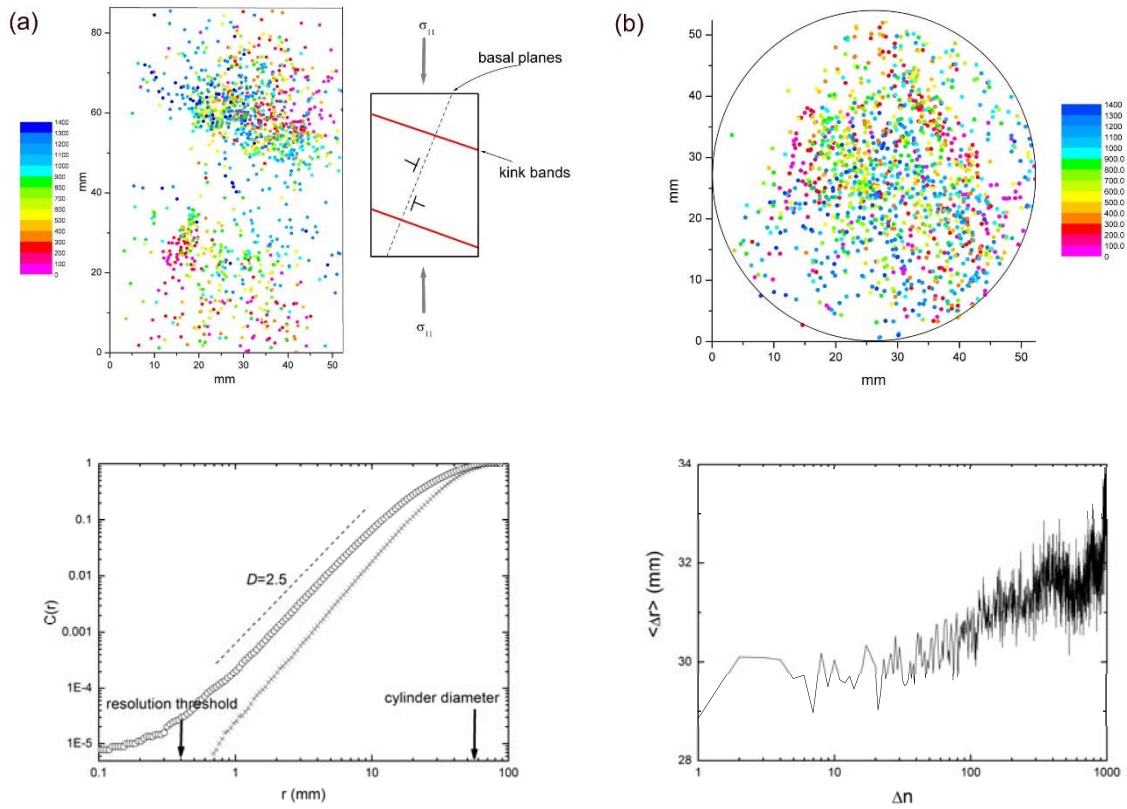

*Fig. 5.* Spatial clustering and space/time coupling of dislocation avalanches: (a) vertical and (b) horizontal projections of the hypocenters of a 3D mapping of dislocation avalanches during a creep tests on an ice single crystal. The hypocenters have been colored as a function of their occurrence rank within the sequence of events. (c) Correlation integral analysis of the hypocenters locations (circles). $C(r)$, which is the probability of two locations being separated by less than $r$, scales as $C(r) \sim r^D$, with $D = 2.5 \pm 0.1$. A similar analysis performed for the same number of randomly distributed locations (crosses) gives $D = 2.9 \pm 0.1$, in agreement with a dense structure ($D = 3$) (From [69]). (d) Evolution of the average distance $\langle \Delta r \rangle$ between two hypocenters separated in time by $\Delta n$ events, illustrating the space/time coupling of the collective dynamics.

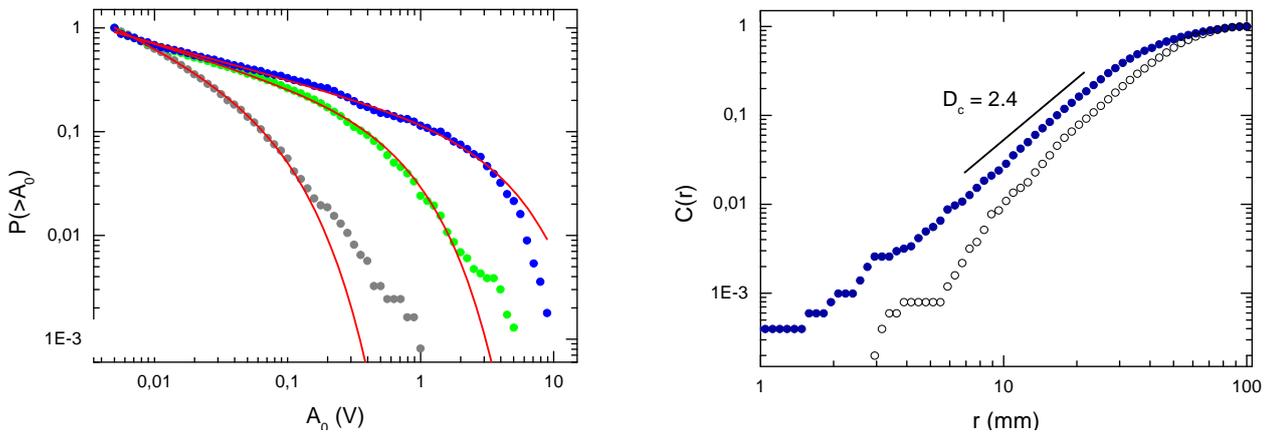

*Fig. 6.* Dislocation avalanches in ice polycrystals: (a) Cumulative probability functions (CDF) of the acoustic burst amplitudes ($A_{max}$) recorded during the compression creep of ice polycrystals at -10°C. From left to right: $\langle d \rangle = 260$ µm, $\sigma_I$=0.67 MPa; $\langle d \rangle = 1.05$ mm, $\sigma_I$=0.57 MPa; $\langle d \rangle = 1.92$ mm, $\sigma_I$=0.54 MPa. The CDFs, instead of the PDFs, have been shown to exemplify the upper cut-offs. The thin solid lines are best fits with the empirical relation $P(> A_{max}) \sim A_{max}^{-\kappa_A+1} \exp(\frac{-A_{max}}{A_c})$. (b) Same as Fig. 5c for a compression test (-10°C, $\sigma_I$=0.81 MPa) performed on a polycrystal with $\langle d \rangle = 2.59$ mm. The scaling regime ($D = 2.4 \pm 0.1$) extends far beyond the grain size (blue plain circles). Open circles: similar analysis for the same number of randomly distributed locations.

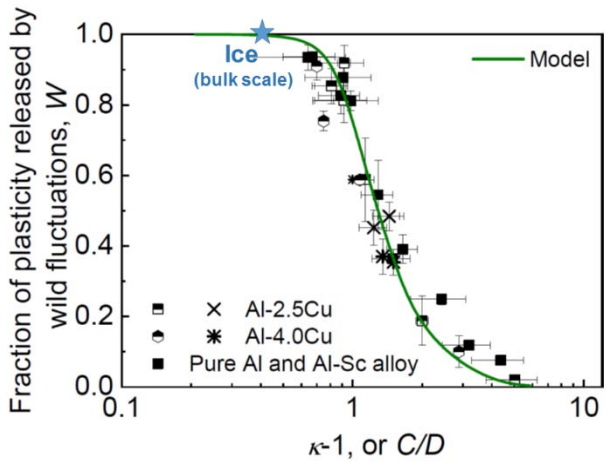

*Fig. 7.* Relationship between the fraction of plasticity released by wild fluctuations, *W*, and the associated exponent of the power law tail of the distribution of avalanche sizes, for micropillars experiments on Al and various Al-alloys (pillar diameters varying from 500 to 3500 nm), and for ice at bulk scale (adapted from [37]). For micropillars, avalanche sizes and wildness have been measured

directly from the stress-strain curves, but from AE data in ice. The green solid line represents the prediction given by the stochastic model presented in [17] and [37]. In this representation, ice appears as an extreme case of wildness.

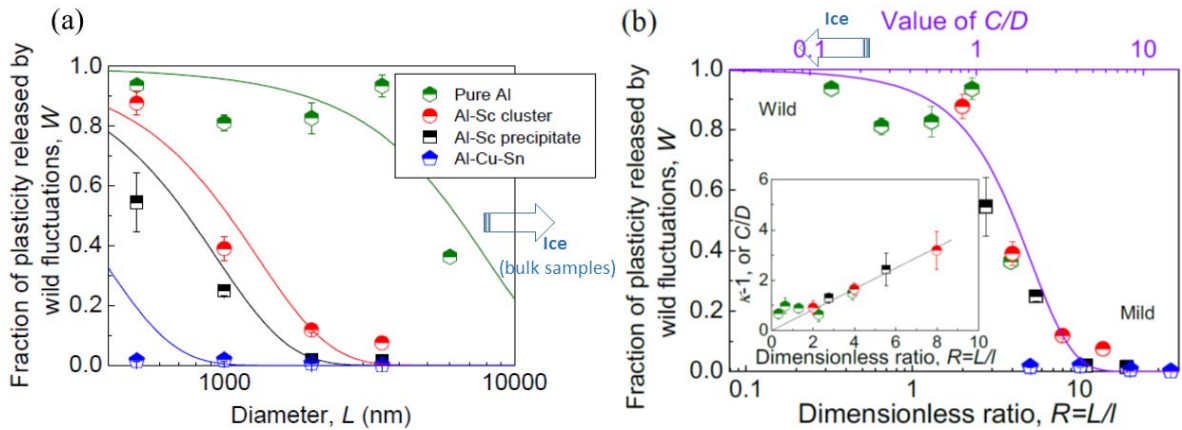

*Fig. 8.* (a) Wildness $W$ as a function of micropillar size $L$ for pure Al and various Al-alloys. (b) When renormalized by the internal length scale $l$ (see text for details), the same data collapse onto a single "universal" curve. The position of ice macroscopic samples in these diagrams has been sketched (adapted from [37]).